# Barium Calcium Zirconium Titanate Thin Film-Based Capacitive Thermoelectric Converter for Low-Grade Waste Heat

Mohammad K. Al Thehaiban[1]*, Vladimir S. Getov[2], Qiaomu Yao[1], Chukwudike C. Ukeje[1] and Peter K. Petrov[1]

[1]Department of Materials, Imperial College London, Exhibition Rd, South Kensington, London SW7 2AZ, UK

[2]School of Computer Science and Engineering, University of Westminster, 115 New Cavendish St, London W1W 6UW, UK

*m.al-thehaiban19@imperial.ac.uk

## ABSTRACT

A capacitive thermoelectric device can harvest thermal energy and convert it to electrical energy by employing a temperature-dependent dielectric material whose permittivity sharply changes with temperature. Electricity can be generated by fluctuating the temperature of the capacitor. Currently, capacitive thermoelectric devices are not broadly used, which can be attributed to the low efficiency of the existing solutions, the lack of dielectric materials with suitable temperature non-linearity of the dielectric permittivity, and the complexity of modulating heat flux on the dielectric material. Here, we propose a device based on $(Ba_{0.85}Ca_{0.15})(Ti_{0.92}Zr_{0.08})O_3$ and $(Ba_{0.73}Ca_{0.27})(Ti_{0.98}Zr_{0.02})O_3$ thin films. The estimated power output under different operation conditions and dynamic workload of an Intel E5-2630 microprocessor show that these thin film materials are promising and can potentially be used for a capacitive thermoelectric converter.

## Introduction

Waste heat is classified into high, medium, and low grades based on the temperature range (see Table 1). 50% of the total waste heat falls in the low-grade category, which is the most challenging to recover compared to medium-grade or high-grade heat[1-4]. The heat emitted from electronic devices is considered low-grade waste heat. For example, a typical cell phone gives off around 0.2 W of heat when idle and closer to 1 W while making calls or running processor-intensive applications[5]. The heat generated by a tablet or laptop is much higher. Almost all of this heat is never recovered; therefore, focusing on the thermal characteristics of smaller mobile devices is a promising future development, especially since the number of mobile devices is expected to reach 18.22 billion worldwide by 2025, an increase of 4.2 billion devices compared to 2020 levels[6].

**Table 1.** Emitted heat sources

| Grade | Temperature range | Typical Source |
|---|---|---|
| High | > 650 ºC | Directly combustion process |
| Medium | 230 ºC- 650 ºC | Exhaust of combustion units |
| Low | < 230 ºC | products and the equipment of process units |

At present, the solid-state devices used for capturing low-grade waste heat rely on the inherent properties of the materials employed. These materials can generate electricity from heat through thermoelectric (Seebeck effect), pyroelectric (spontaneous polarisation), or capacitive (dielectric permittivity) processes. However, the main obstacle arises from the fact that low-grade heat harvesting happens with very low efficiency. One potential solution is instead of enhancing the conversion efficiency, to increase the conversion rate by employing thin film-based devices.

A thermoelectric generator (TEG) is a solid-state device that directly converts temperature gradient into electricity through the Seebeck or thermoelectric effect. A thermoelectric device consists of two dissimilar thermoelectric materials joined at their respective ends by an *n*-type and a p-type semiconductor. The two thermoelectric materials must be thermally parallel and in electrical series[7,8].

The conversion efficiency of a thermoelectric device mainly depends on the materials' figure of merit[9,10].

$$ZT = \frac{S^2 \sigma T}{\kappa} \qquad (1)$$

Where *S*, *σ*, *T* and *κ* are Seebeck coefficient, electrical conductivity, absolute temperature, and thermal conductivity, respectively. Simultaneously attaining good electrical properties and low thermal conductivity will lead to a high *zT* value. Most of the impressive advancements in zT have been achieved by effectively decreasing thermal conductivity through the enhancement of phonon scattering in material structures. This includes mechanisms such as dislocations, grain boundaries and interfaces. [1,11-13].

Yang et al.[14] conducted a study in which a thin-film thermoelectric generator was created using $Sb_2Te_3$. They grew the thin film on 1mm thick $Pb(Zr_{0.54}Ti_{0.46})O_3$ ceramic sheets using magnetron sputtering. Their thermoelectric generator achieved output power and output power density of 342.12 nW and 2.22 mW/cm$^2$, respectively at ΔT = 42 K. In the work of Karthikeyan et al.[15] 100nm of both p-type tin telluride(SnTe) and n-type lead telluride (PbTe) thin films were fabricated using thermal evaporation grown onto a flexible polyimide substrate. The thermoelectric generator is based on SnTe–PbTe, consisting of 4 p-n pairs interconnected by a 50 nm thick aluminium film. Their results showed that the fabricated TEG produced 8.5 mW/cm$^2$ at a temperature difference of ΔT = 120 °C. Ren et al.[16] utilised a thermal evaporation process to create 14 thermoelectric (TE) chips on a polyimide film. Each thermoelectric chip was constructed by alternately depositing 4 pairs of thin films of $Bi_{0.5}Sb_{1.5}Te_3$ (*p*-type) and $Bi_2Te_{2.8}Se_{0.3}$ (*n*-type) onto the polyimide film using thermal evaporation. The results showed that for a 93 °C temperature gradient, the fabricated thermoelectric chips generated an output power of 18.625 μW/cm$^2$.

Pyroelectric and capacitive thermoelectric devices can both convert heat fluctuation to electrical energy. They use dielectric materials with specific crystal structures that exhibit a unique axis of symmetry and lack a centre of symmetry; these are referred to as "polar" materials[17-21]. These materials show a temperature and electrical field dependence on the spontaneous polarisation and the dielectric permittivity.

A pyroelectric device operation is based on the material's spontaneous polarisation caused by temperature change, which leads to variations in its surface charge density. The pyroelectric coefficient, denoted as p(T), represents the rate of change of spontaneous polarisation in the material with respect to temperature without the influence of an applied electric field or stress[22,23].

$$p(T) = d\,P_S/dT \qquad (2)$$

The electric current $i_p$ generated during the heat fluctuation ($\frac{dT}{dt}$) of a pyroelectric device connected to an external circuit with electrode surface area A can be written as follows:

$$i_p = A \, p(T) \frac{dT}{dt} \quad (3)$$

Ashwath Aravindhan reported lead scandium tantalate, $Pb(Sc_{1/2}Ta_{1/2})O_3$ (PST) thin film for pyroelectric energy conversion[24], where it is phase transition temperature is around room temperature. The fabricated thin film with a thickness of 165 nm was grown on a c-sapphire substrate using a chemical solution deposition. The maximum energy density achieved is 9.1 J/cm³ per cycle when there is a temperature change ($\Delta T$) of 150 K. Bhatia[25] fabricated a 200 nm thick $BaTiO_3$ thin-film capacitor fabricated using pulsed laser deposition with top and bottom epitaxial $SrRuO_3$ electrodes, grown on a $GdScO_3$ (110) single crystal substrate. They have achieved a maximum power density of 30 W/cm³ working at a temperature range of 20–120°C.

Pandya[26] fabricated 150 nm $0.68Pb(Mg_{1/3}Nb_{2/3})O_3$–$0.32PbTiO_3$ (PMN–PT)/20 nm $Ba_{0.5}Sr_{0.5}RuO_3$ heterostructures by pulsed laser deposition grown on (110)-oriented, single-crystalline $NdScO_3$ substrate. Their results showed a maximum energy density of 1.06 J/cm³ at a temperature difference of $\Delta T = 10$ K.

For a capacitive thermoelectric device, the value of the polar dielectric material's permittivity changes with temperature. This will vary the capacitance value and, as a result, the electrical energy stored in the capacitor[27-30]. The fundamental concept behind converting thermal to electric energy (W) stored in a capacitor with capacitance (C) for a capacitive thermoelectric device is as follows:

$$W = \frac{Q^2}{2C} \quad (4)$$

Where Q is the electrical charge in the capacitor, which increases because of thermal energy when the capacitance value decreases, assuming the charge of the capacitor stays constant. The operational principle of capacitive converters based on the Clingman thermodynamic cycle is illustrated in Figure 1. Initially, at point 1, starting at a temperature $T_1$, the capacitor is charged using an external source. After that, the switch is opened at point 2, and the system is gradually heated to reach temperature $T_2$, while the electric charge remains constant. At the same time the temperature-dependent permittivity of the dielectric material changes, which causes the increase of the potential in the capacitor from $V_1$ to $V_2$. When the switch is moved to position 3, the voltage built in the capacitor pushes the electrical charges through the external circuit (e.g., the load resistor).

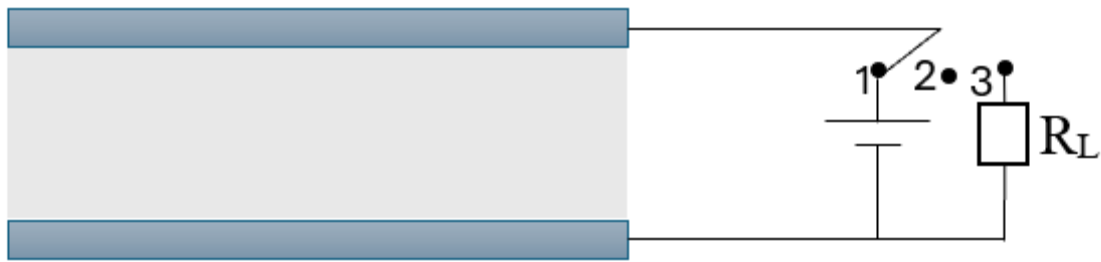

**Figure 1** Schematic diagram of a capacitor with non-linear dielectric material with temperature

After the capacitor has been discharged, the system undergoes a cooling process to reach temperature $T_1$, when the capacitor is charged again, and the cycle is repeated. Although pyroelectric devices and capacitive thermoelectric converters share similarities, the distinction lies in the way the change of the temperature is utilised. In a pyroelectric device, the temperature change is used to generate electrical charges (Olsen thermodynamic cycle), while in a capacitive TE converter, it is used to push the charges stored in the capacitor (Clingman thermodynamic cycle). Further details about Clingman and Olsen thermodynamic cycles can be found elsewhere[31].

The first mention of utilising the non-linear dielectric behaviour with temperature to convert energy was in 1961[27]. W. H. Clingman and R.G. Moore, Jr. proposed a design for a ferroelectric converter circuit based on barium titanate. They have found that power output can be increased by adjusting the temperature difference and electrical field strength. Their findings suggest that these converters can be used in applications that require lightweight devices with a periodic heat source for operation, such as a spinning space vehicle that undergoes repeated cycles of heating and cooling as a result of regular exposure to solar radiation, providing an optimistic vision for the potential of energy conversion in the years to come.

A year later, Childress [28] estimated a power-generating capacity of 0.1 cm thick barium titanate with a 30°C temperature difference. They calculated a power-generating capacity of roughly 32 W/lb (70.6 W/kg) per temperature cycle, per sec. In their analysis, the periodic heat transfer of the dielectric layer is a serious limitation for converting thermal energy.

Both of these early studies on barium titanate ceramics showed that achieving fast heating-cooling cycling of such a device is challenging. However, using a material with non-linear temperature dependence in thin film form will significantly increase the speed of the heating-cooling cycle and generate a useful amount of power based on the increased conversion rate rather than efficiency. It is worth mentioning that a capacitive thermoelectrical device is electrically identical to a dielectric bolometer, which converts thermal to electrical energy for IR sensing[32,33]. To the best of our knowledge, the available data in the literature regarding thin film-based capacitive TE converters is based on modelling. V.A. Volpyas et al.[29] made a thermodynamic analysis of a capacitive thermoelectric converter based on barium strontium titanate ferroelectric film incorporated into a metal–insulator–metal (MIM) structure. Their device comprises a heat flux modulator, heater, and thin film capacitor on a cooled substrate surface. They have further evaluated the generated electrical power as the permittivity of the MIM changes under periodic heating and cooling cycles. This study has found that the electrical power generated is governed by the steepness of permittivity dependence on the temperature and the thermocycling frequency. Ferroelectric materials exhibit spontaneous electric polarisation that can be reversed by applying an external electric field within a temperature range.

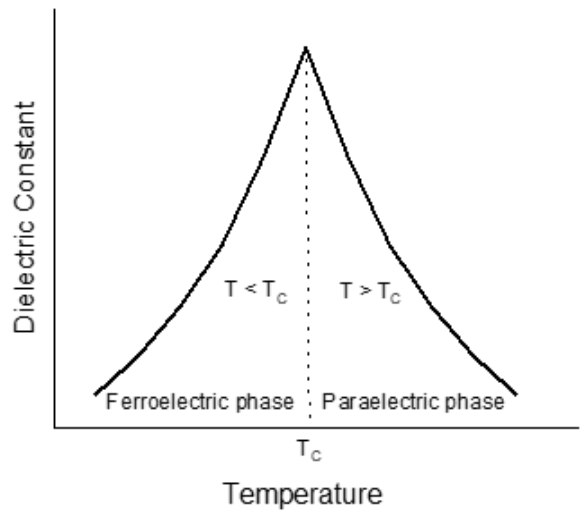

**Figure 2** . Illustration of the phase changes in a ferroelectric material around Curie temperature ($T_C$)

These materials (see Figure 2) have a phase transition temperature from the ferroelectric phase (with the presence of spontaneous polarisation) to the paraelectric phase (absence of spontaneous polarisation). The phase transition temperature is known as Curie temperature. Ferroelectric materials having a first-order phase transition show a sharp change of permittivity with temperature where the maximum permittivity value is at Curie temperature. Also, dielectric loss in the paraelectric phase is lower in comparison to the dielectric loss in the ferroelectric phase where the stress-strain resulting from extrusions in the domain walls will convert electrical energy into heat[34,35].

Choosing a ferroelectric material with a sharp change of permittivity with a temperature operating at the paraelectric phase will increase the harvested energy and optimise the heating-cooling cycle.

The research result achieved by Liu et al. on the lead-free barium calcium zirconium titanate (BCZT) ferroelectric system is quite attractive since it offers an alternative to PZT-based systems. According to their findings, this new system exhibits a remarkably high piezoelectric coefficient and dielectric constant, which could benefit various applications[36]. Their method for achieving a high piezoelectric coefficient and a non-linear dielectric behaviour with temperature involves balancing the materials' stoichiometry near a composition-induced phase transition between two ferroelectric phases. This transition is referred to as the 'morphotropic phase boundary' or MPB [36]. Ferroelectric compositions near morphotropic phase boundaries (MPBs) exhibit exceptional properties that surpass those found in compositions situated farther from these boundaries. Additionally, the ferro-para transition induced by composition at MPBs can destabilise polarisation states, allowing for easy rotation of polarisation direction through external stress or electric field[37]. Consequently, it will result in high piezoelectricity and permittivity [38]. This resulted in a rise of investigations of BCZT in thin film form for applications in piezoelectric[39], optical-thermal sensors[40], and magneto-dielectric devices[41]. Nevertheless, the temperature-dependent non-linear dielectric properties have yet to be explored for capacitive thermoelectric device applications.

Almost all of today's electronic devices are built using semiconductor materials, which has been the case since the middle of the 20$^{th}$ century[42].

Although the variety of materials keeps widening, silicon-based technologies remain by far the dominating choice for current and future electronic devices. At present, the most important application domains include Internet-of-things edge (IoTe) devices, cyber-physical systems, smartphones, personal augmentation devices, large-scale computers, and distributed systems [43].

Normally, silicon-based semiconductor components are designed as low-power electronic devices with supply voltage ranging from 0.6 V to less than 20 V. For example, one of the most popular current solutions – the Field-Effect Transistor (FET) technology – operates with a voltage of 0.7 V at the device level, projected to trend down slowly to 0.6 V over the next decade[44].

By contrast, the supply voltage for the traditional Complementary Metal Oxide Semiconductor (CMOS) technology is the only one that goes beyond 6 V, covering the range from 3 to 18 V. However, operation above 15 V is not recommended because of high dynamic power consumption and the risk of noise spikes on the power supply exceeding the breakdown voltage of around 20 V[45]. Therefore, the supply voltage range for our analysis was selected between 0.5 V and 15 V.

In this paper, we have estimated the power output of a capacitive thermoelectric device based on thin films with stoichiometry located near the phase boundary $(Ba_{0.73}Ca_{0.27})(Ti_{0.98}Zr_{0.02})O_3$, referred further as $Ba_{0.73}$ and $(Ba_{0.85}Ca_{0.15})(Ti_{0.92}Zr_{0.08})O_3$, referred further as $Ba_{0.85}$ and have assessed their suitability to be used to harvest low-grade waste heat.

## Results and discussion

### Thin film performance

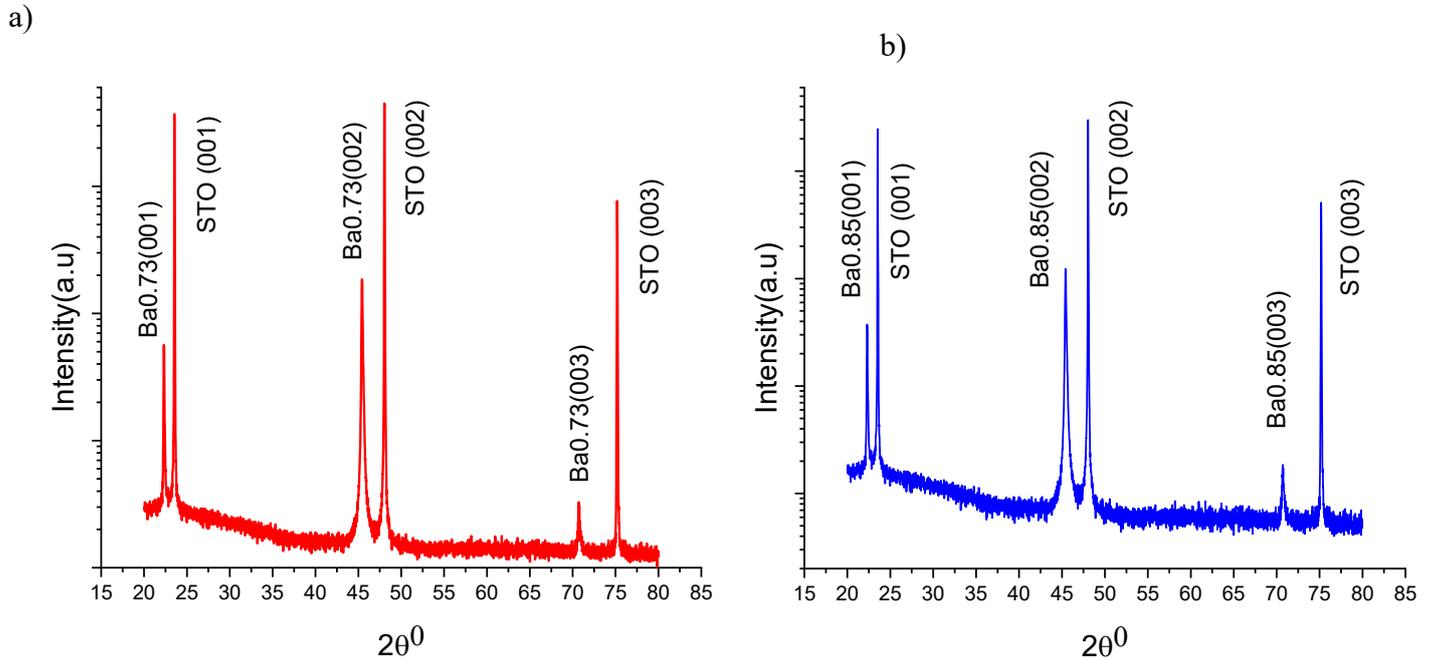

**Figure 3 .** XRD pattern of a) Ba0.73 and b) Ba0.85 thin films, respectively. They confirm that both thin films are grown on STO substrate epitaxially with sole (00*l*) orientation.

Figures 3 (a) and (b) show the XRD pattern of thin films $Ba_{0.73}$ and $Ba_{0.85}$; respectively. They confirm that both thin films are grown on STO substrate epitaxially with sole (00*l*) orientation. Table 2 lists the lattice parameters and residual strain of each thin film. Both films show relatively low residual strain, which is a result of the deposition condition and the low lattice misfit.

*Table 2 .* Lattice parameters and in-plane and out-of-plane strain for epitaxially grown thin films

| Composition | Lattice parameters (Å) | | Strain | |
|---|---|---|---|---|
| | a | c | in-plane  a% | out-of-plane c% |
| $Ba_{0.73}$ | 3.960 | 4.008 | -0.65 | 0.72 |
| $Ba_{0.85}$ | 3.990 | 4.040 | -0.19 | 0.69 |

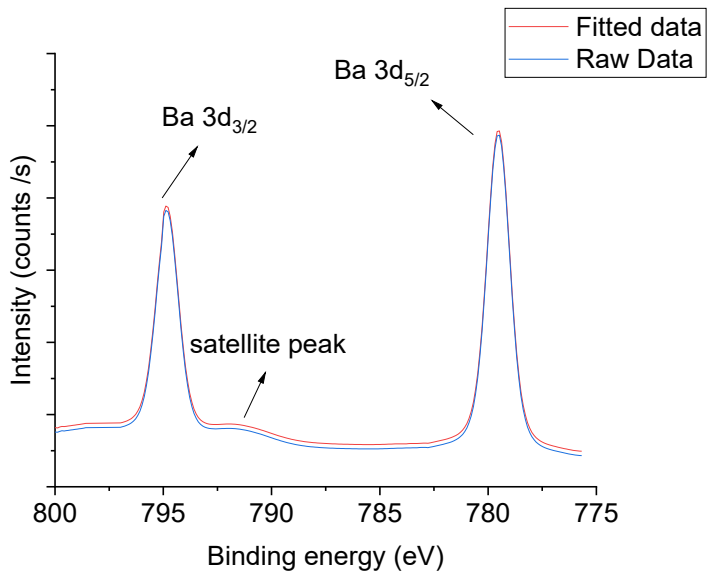
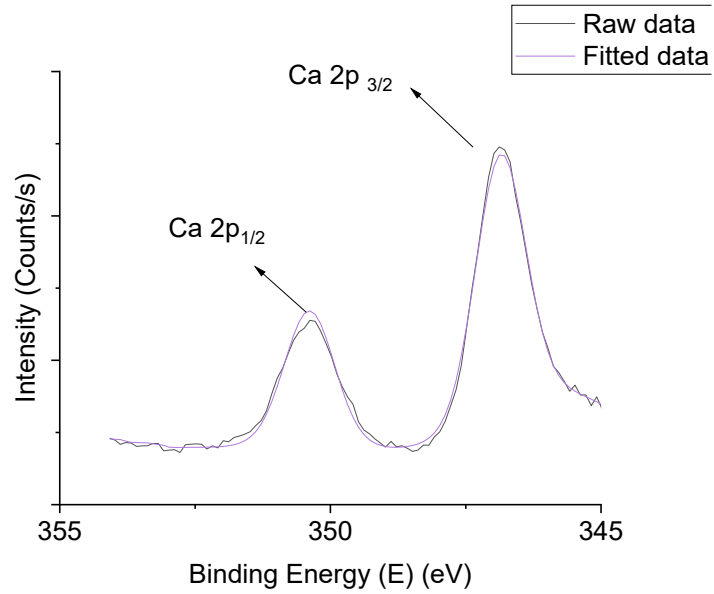
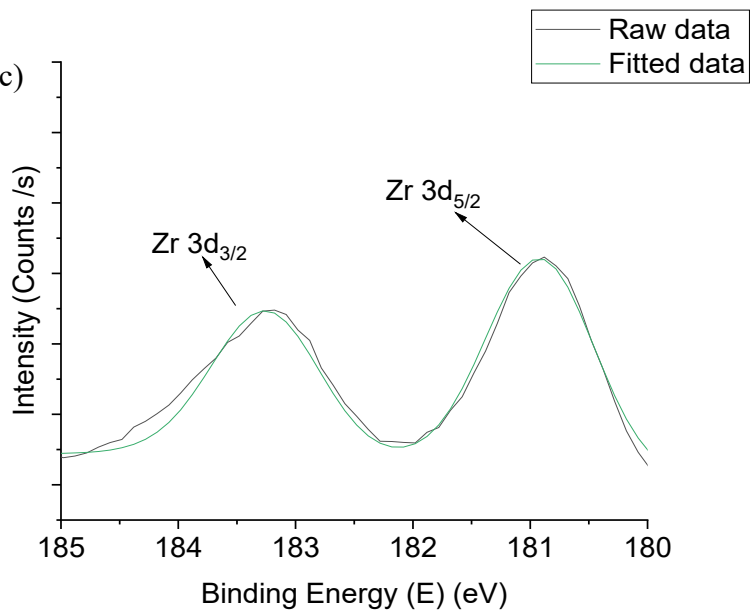
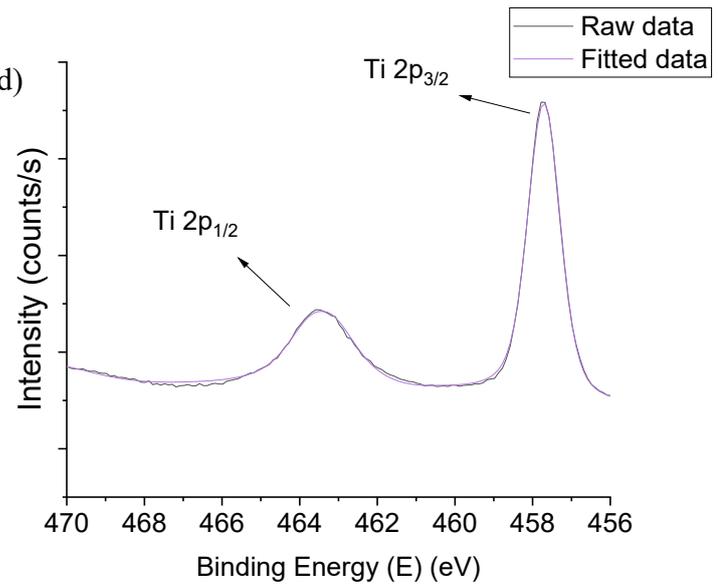

**Figure 4**. XPS spectra fitting results of Ba, Ca, Zr and Ti atoms of Ba0.73 thin film. (a) Ba3d XPS spectra region, b) Ca2p XPS spectra region, c) Zr3d spectra region and d)Ti2p XPS spectra region.

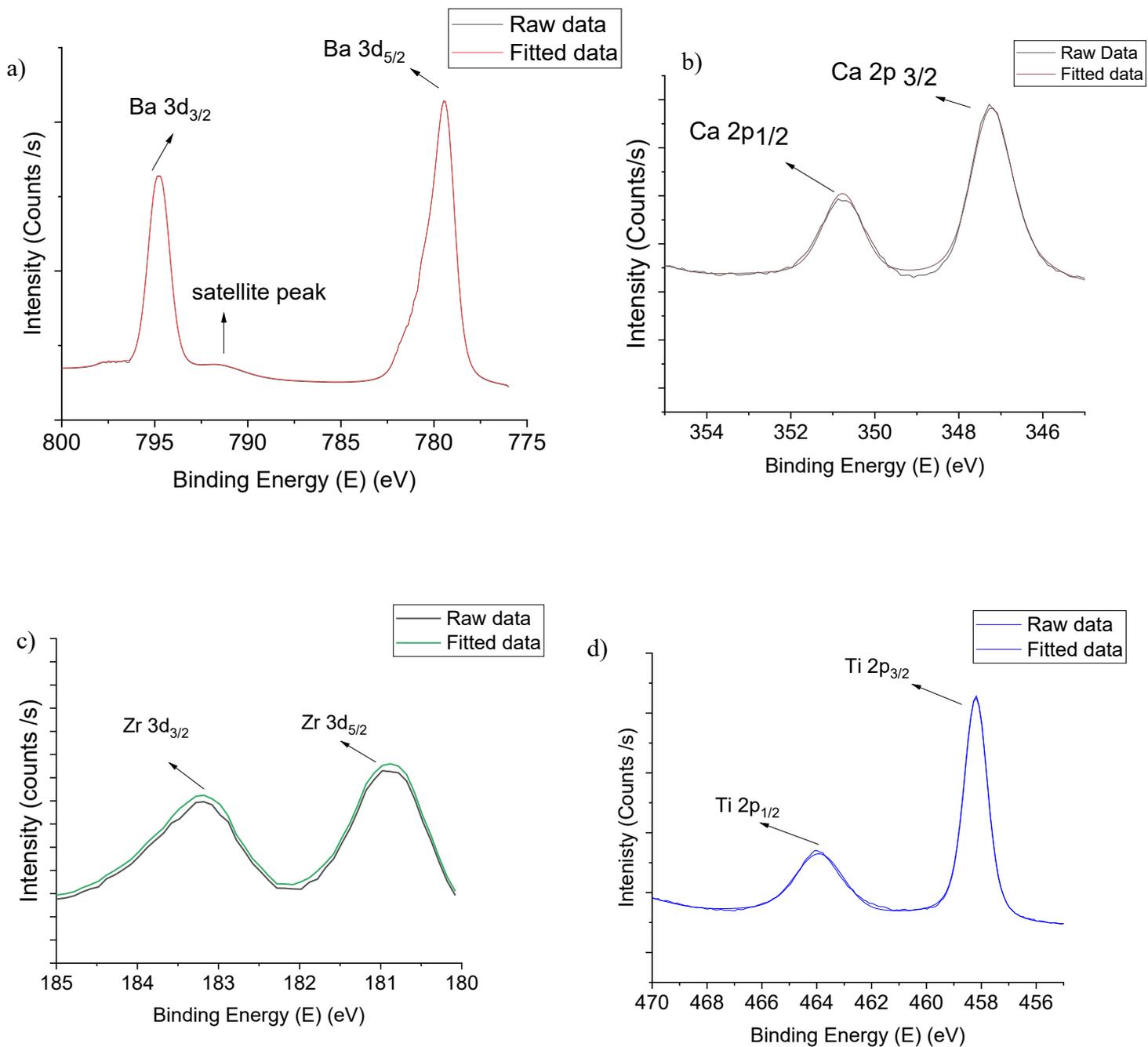

**Figure 5.** XPS spectra fitting results of Ba, Ca, Zr and Ti atoms of Ba0.73 thin film. (a) Ba3d XPS spectra region, b) Ca2p XPS spectra region, c) Zr3d spectra region and d)Ti2p XPS spectra region.

X-ray photoelectron spectroscopy (XPS) was used to characterise the BCZT thin films and confirm their stoichiometry. Figures 4 and 5 show peaks corresponding to photoemissions from the core-level of XPS spectra for Ba 3d, Ti 2p, Ca 2p, and Zr 3d. These peaks are referenced in the literature to validate the chemical composition of BCZT thin films.[46-48]. Based on the XPS spectra, the atomic concentrations were calculated and the results relevant to Ba, Ca, Zr and Ti are presented in Tables 3 and 4 for both thin films. Also, the XPS spectra were used to calculate the cationic site occupancy ratio (B/A site) for the perovskite ABO3-type BCZT unit cell. It was found to be (Ti+Zr/Ba+Ca) =0.97 for $Ba_{0.73}$ thin film and 0.98 for $Ba_{0.85}$ thin film. These results for both thin films confirm that the chemical composition is as expected.

**Table 3 .** Atomic % of elements of sample Ba0.73

| Element | Expected | Measured |
|---|---|---|
| Ba | 0.73 | 0.74 |
| Ca | 0.27 | 0.28 |
| Zr | 0.02 | 0.03 |
| Ti | 0.98 | 0.96 |

**Table 4 .** Atomic % of elements of sample Ba0.85

| Element | Expected | Measured |
|---|---|---|
| Ba | 0.85 | 0.84 |
| Ca | 0.15 | 0.18 |
| Zr | 0.08 | 0.09 |
| Ti | 0.92 | 0.91 |

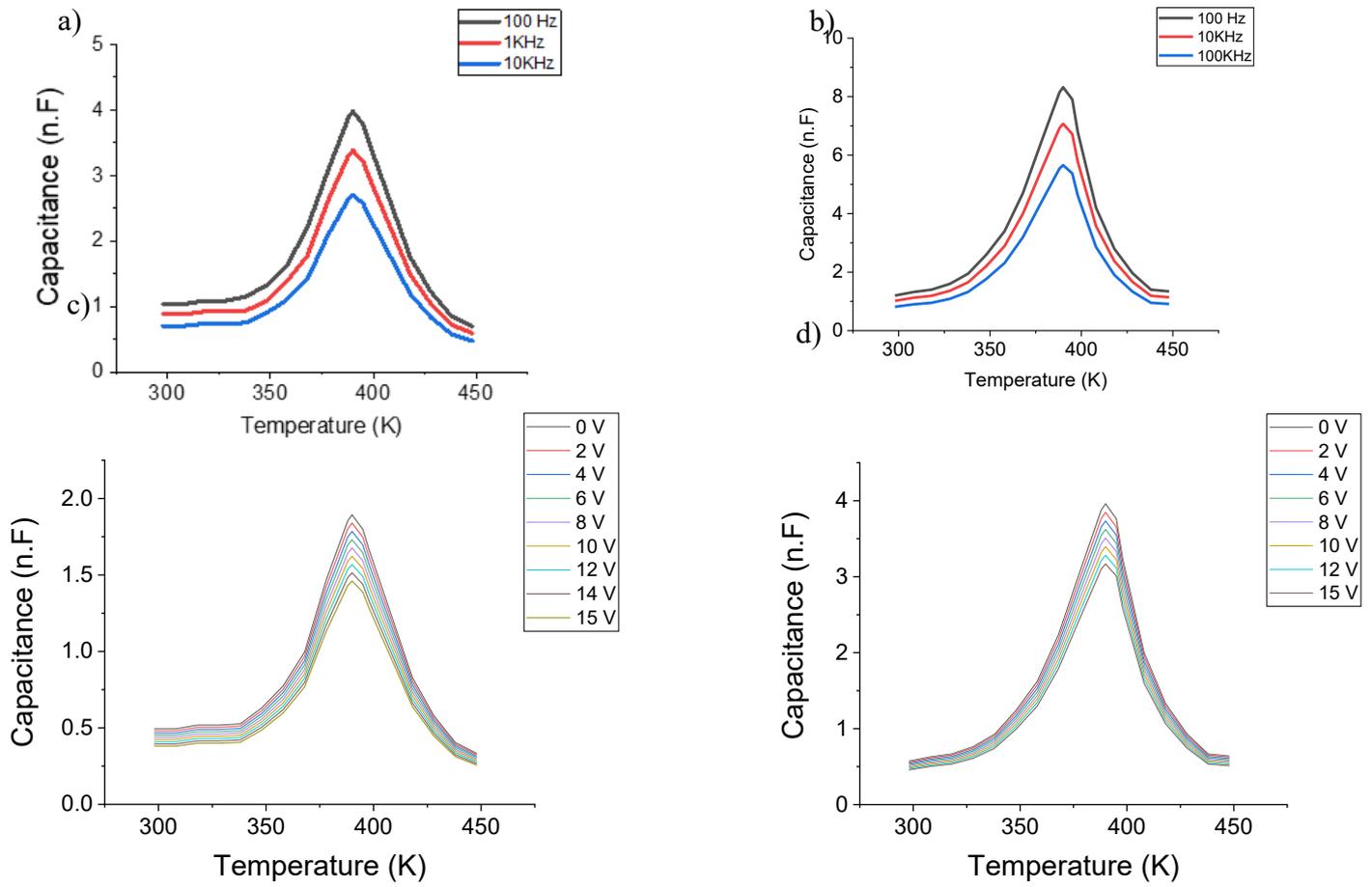

**Figure 6.** Capacitance as a function of temperature at various frequencies for samples a) Ba0.73, b)Ba0.85 and at various DC bias voltage for samples (c) Ba0.73, d) Ba0.85. Both samples show similar Curie temperatures and a capacitance that sharply changes with the temperature.

Figure 6 (a) and (b) show the capacitance as a function of the temperature applying AC voltage while, Figure 6 (c) and (d) show the DC bias voltage measured for samples made using both types of thin films. They both show a capacitance that sharply changes with the temperature with a maximum (corresponding to the Curie temperature of the ferroelectric material) at similar temperatures. The capacitance values for samples $Ba_{0.73}$ and $Ba_{0.85}$ change around four and seven folds, respectively. These results indicate that these films are good candidates for thin film-based capacitive thermoelectric converters.

**Estimated power output**

The power generated by a capacitive TE converter can be estimated as follows [30]:

$$P = 0.5 \cdot (k - 1) \cdot C_{max} \cdot V^2 \cdot f \quad (5)$$

Where $C$ is the capacitance, $C_{max}$ is the maximum capacitance value, $C_{min}$ is the minimum capacitance value, $k = C_{max}/C_{min}$, $V$ is bias voltage, and $f$ is the frequency of temperature cycling.

a) 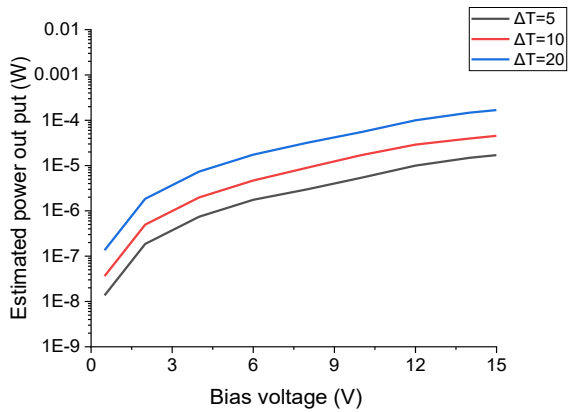

b) 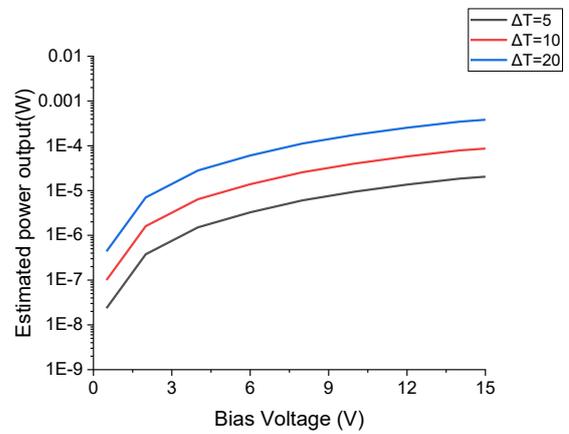

**Figure 7 .** The estimated output power of a thermoelectric converter operating with a ΔT=5, 10 and 20, as a function of applied bias voltage for samples a) Ba0.73, and b) Ba0.85

Figure 7 displays the calculated power output of the $Ba_{0.73}$ and $Ba_{0.85}$ thin film-based devices, operating with a frequency of temperature cycling of 2 kHz, 15 V bias and various temperature changes (ΔT). The amount of power output of $Ba_{0.85}$ is higher than $Ba_{0.73}$ at all ΔT and frequency of heating and cooling. Also, it is worth noting that the amount of converted energy can be increased by operating in a wider temperature range and/or by increasing the bias voltage. Taking an average microprocessor such as Intel E5-2630 as an example, one can observe the thermal characteristics (shown in Figure 7) depending on the artificially generated workload by the Firestarter software[49]. This software is specially designed to stress a processor to its fullest and provides opportunities to select partial or full workloads while using certain numbers of cores with its 'stress code' utility.

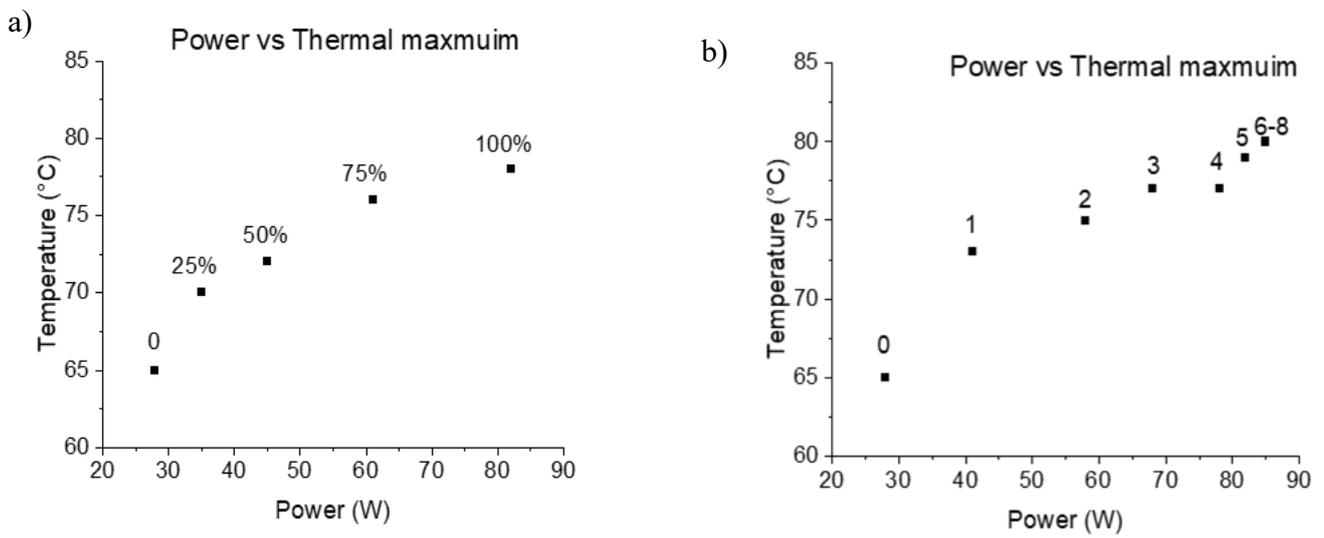

The results presented in Figure 8 show that a deviation in temperature in the range of 15°C is possible for higher workloads. However, this requires a longer period, while a deviation in temperature in the range of 5°C is much faster at the beginning of the thermal rise or fall due to the exponential change of the thermal energy[50].

**Figure 8.** Temperature of an Intel E5-2630 microprocessor as a function of the power consumption. The numbers in the graphs represent (a) the workload in % and (b) the number of active cores.

**Conclusion**

Thus, two barium calcium zirconium titanate-based thin films were considered for the development of a capacitive TE converter. The estimated power output of a device operating with a $\Delta T = 20°C$, at a bias voltage of 15 V ranged from 0.1 mW (the equivalent of 0.5 mW/mm²) to 0.3 mW (the equivalent of 1.5 mW/mm²), which is significantly higher for that temperature range compared to the previously reported values[14-16,24-26]. An Intel E5-2630 microprocessor with dynamic workload was considered as a heat-modulated source. The possibility of achieving a suitable thermoelectric conversion was demonstrated given an intelligent control of the processor's workload, which does not compromise its performance.

It is also worth noting the double benefit of the proposed thermoelectric converter. In addition to the harvested energy, it will contribute substantially to the cooling of the electronic devices, hence, reducing the demand of external energy required for their active cooling.

## Methods

### Thin Films Design and Preparations

The thin films were deposited via the pulsed laser deposition (PLD) method using in-house sintered ceramic pucks with the desired stoichiometry. Two types of films with a thickness of 300 nm were deposited onto strontium titanate (STO) substrates (Crystal 1STO 105E, 5 x 5 mm$^2$, (100) oriented, one side polished). The substrates were secured by silver paste on a stainless resistive heater. The optimised deposition conditions for Ba$_{0.73}$ thin film fabricated at 730℃, 100mTorr oxygen partial pressure, and laser fluence 2 J/cm$^2$. Ba$_{0.85}$ thin film was fabricated at 780℃, 300mTorr oxygen partial pressure and laser fluence 2 J/cm$^2$. The thickness of both films was measured using a (KLA) Tencor D-600 profiler. The crystal orientation of the film was analysed using X-ray diffraction (XRD) (Empyrean multipurpose diffractometer, Malvern Panalytical). A Thermo-Scientific K-alpha+ X-ray photoelectron spectrometer investigated the stoichiometry of the thin films.

To facilitate electrical measurement on thin films, electrode layers of Au buffered with TiO$_x$ were deposited by magnetron sputtering (HEX, Korvus Technology). After the electrode deposition, the samples were subjected to photolithography using an OAI model 200 mask aligner and ion milling using the scia Mill 150 system. The patterned capacitor structures had electrodes with a length and width of 450 μm and a 10 μm gap between them. Thin film dielectric properties were measured using an LCR meter (Hewlett Packard 4263B) attached to a probe station (Janis Inc.) in a temperature range 298-450 K at frequencies of 100 Hz, 1kHz, 10kHz and 1MHz under DC bias ranging from 0 V up to 15 V.

### Electronic Device Selection as Modulated Heat Source

In our work, the Intel E5-2630 microprocessor with varied workloads was considered as the source of modulated heat flux. This microprocessor is a mid-range homogeneous multi-core Intel Haswell device with 8 cores, which provide dedicated resources for the execution of individual threads. In contrast, the remaining microprocessor components are shared between all threads, and therefore, the workload depends on the balance between the number of active cores and the shared components.

**References**

[1] Kishore, R. & Priya, S. A Review on Low-Grade Thermal Energy Harvesting: Materials, Methods and Devices. *Materials* **11**, 1433 (2018).

[2] Johnson, I., Choate, W. T. & Davidson, A. Waste Heat Recovery. Technology and Opportunities in U.S. Industry. *BCS Inc*; https://www.osti.gov/biblio/1218716 (2008)

[3] Vance, D. *et al.* Estimation of and barriers to waste heat recovery from harsh environments in industrial processes. *Journal of Cleaner Production* **222**, (2019).

[4] Erickson, D. C., Anand, G., & Kyung, I. Heat-Activated Dual-Function Absorption Cycle. *ASHRAE Transactions* **110**. (2004)

[5] Carroll, A., & Heiser, G., *An Analysis of Power Consumption in a Smartphone*, USENIX Annual Technical Conference, pp. 21–21, USENIX Association, 2010

[6] Laricchia, F. Number of mobile devices worldwide 2020-2025. *Statista* https://www.statista.com/statistics/245501/multiple-mobile-device-ownership-worldwide/ (2021)

[7] Polozine, A., Sirotinskaya, S. & Schaeffer, L. History of development of thermoelectric materials for electric power generation and criteria of their quality. *Materials Research* **17**, 1260–1267 (2014).

[8] Shi, X.-L., Zou, J. & Chen, Z.-G. Advanced Thermoelectric Design: From Materials and Structures to Devices. *Chemical Reviews* **120**, 7399–7515 (2020).

[9] Kim, H. S., Liu, W., Chen, G., Chu, C.-W. & Ren, Z. Relationship between thermoelectric figure of merit and energy conversion efficiency. *Proceedings of the National Academy of Sciences* **112**, 8205–8210 (2015).

[10] David Michael Rowe & Ebooks Corporation. *Thermoelectrics handbook: macro to nano-structured materials*. (Taylor & Francis, 2006).

[11] Rowe, D. M. *CRC Handbook of Thermoelectrics*. (CRC Press, 2018).

[12] Biswas, K. *et al.* High-performance bulk thermoelectrics with all-scale hierarchical architectures. *Nature* **489**, 414–418 (2012).

[13] He, J., Girard, S. N., Kanatzidis, M. G. & Dravid, V. P. Microstructure-Lattice Thermal Conductivity Correlation in Nanostructured PbTe$_{0.7}$S$_{0.3}$ Thermoelectric Materials. *Advanced Functional Materials* **20**, 764–772 (2010)

[14] Yang, B. *et al.* Regulation of electrical properties via ferroelectric polarization for high performance Sb2Te3 thermoelectric thin films. *Chemical Engineering Journal* **477**, 147005 (2023).

[15] Karthikeyan, V. *et al.* Wearable and flexible thin film thermoelectric module for multi-scale energy harvesting. *Journal of Power Sources* **455**, 227983 (2020).

[16] Ren, W. *et al.* High-performance wearable thermoelectric generator with self-healing, recycling, and Lego-like reconfiguring capabilities. *Science Advances* **7**, eabe0586 (2021).

[17] Thakre, A., Kumar, A., Song, H.-C., Jeong, D.-Y. & Ryu, J. Pyroelectric Energy Conversion and Its Applications—Flexible Energy Harvesters and Sensors. *Sensors* **19**, 2170 (2019).

[18] Wang, Q., Bowen, C. R. & Valev, V. K. Plasmonic-Pyroelectric Materials and Structures. *Advanced Functional Materials* **34**, (2024).

[19] Li, X. *et al.* Pyroelectric and electrocaloric materials. *J. Mater. Chem. C* **1**, 23–37 (2013).

[20] Arli, C., Atilgan, A. R. & Misirlioglu, I. B. Extended contributions to the pyroelectric effect in ferroelectric thin films. *Applied Physics Letters* **124**, (2024).

[21] Olsen, R. B. & Evans, D. Pyroelectric energy conversion: Hysteresis loss and temperature sensitivity of a ferroelectric material. *Journal of Applied Physics* **54**, 5941–5944 (1983).

[22] Sebald, G., Elie Lefeuvre & Guyomar, D. Pyroelectric energy conversion: Optimization principles. *IEEE Transactions on Ultrasonics, Ferroelectrics, and Frequency Control* **55**, 538–551 (2008).

[23] Alpay, S. P., Mantese, J., Trolier-McKinstry, S., Zhang, Q. & Whatmore, R. W. Next-generation electrocaloric and pyroelectric materials for solid-state electrothermal energy interconversion. *MRS Bulletin* **39**, 1099–1111 (2014).

[24] Aravindhan, A. *et al.* Large pyroelectric energy conversion in lead scandium tantalate thin films. *Heliyon* **10**, e30430 (2024).

[25] Bhatia, B., Damodaran, A. R., Cho, H., Martin, L. W. & King, W. P. High-frequency thermal-electrical cycles for pyroelectric energy conversion. *Journal of Applied Physics* **116**, (2014).

[26] Pandya, S. *et al.* Direct Measurement of Pyroelectric and Electrocaloric Effects in Thin Films. *Physical Review Applied* **7**, (2017).

[27] Clingman, W. H. & Moore, R. G. Application of Ferroelectricity to Energy Conversion Processes. *Journal of Applied Physics* **32**, 675–681 (1961).

[28] Childress, J. D. Application of a Ferroelectric Material in an Energy Conversion Device. *Journal of Applied Physics* **33**, 1793–1798 (1962).

[29] Volpyas, V. A., Kozyrev, A. B., O. Soldatenkov & Tepina, E. R. Efficiency of thermoelectric conversion in ferroelectric film capacitive structures. *Technical Physics* **57**, 792–796 (2012).

[30] Kozyrev, A. B., Platonov, R. A. & Soldatenkov, O. I. Thermal-to-electric energy conversion using ferroelectric film capacitors. *Journal of Applied Physics* **116**, (2014).

[31] Pilon, L. & McKinley, I. M. Pyroelectric Energy Conversion. *Annual Review of Heat Transfer* **19,** 279–334 (2016).

[32] Hanel, R. A. Dielectric Bolometer: A New Type of Thermal Radiation Detector*. *Journal of the Optical Society of America* **51**, 220–220 (1961).

[33] Stafsudd, O. M. & Pines, M. Y. Characteristics of KTN Pyroelectric Detectors. *Journal of the Optical Society of America* **62**, 1153 (1972).

[34] Moulson, A. J. & Herbert, J. M. Electroceramics: materials, properties, applications. (Wiley, 2008)

[35] Safa Kasap. Springer Handbook of Electronic and Photonic Materials. (Springer Science & Business Media, 2006).

[36] Liu, W. & Ren, X. Large Piezoelectric Effect in Pb-Free Ceramics. *Physical Review Letters* **103**, (2009).

[37] Uchino, K. *Ferroelectric Devices*. (CRC Press, 2018).

[38] Damjanovic, D. Contributions to the Piezoelectric Effect in Ferroelectric Single Crystals and Ceramics. *Journal of the American Ceramic Society* **88**, 2663–2676 (2005).

[39] Liu, S. *et al.* A flexible and lead-free BCZT thin film nanogenerator for biocompatible energy harvesting. *Materials Chemistry Frontiers* **5,** 4682–4689 (2021).

[40] Kamnoy, M. *et al.* Investigating the thermo-optic properties of BCZT-based temperature sensors. *Materials* **16,** 5202 (2023).

[41] Bhat *et al*. $Ba_{0.85}Ca_{0.15}Zr_{0.1}Ti_{0.90}O_3/CoFe_2O_4/Ba_{0.85}Ca_{0.15}Zr_{0.1}Ti_{0.90}O_3$ Nanoscale Composite Films with 2–2 Connectivity for Magnetoelectric Actuation. *ACS Applied Nano Materials* 5**(12)** (2022).

[42] Chapman, B. Recent U.S. Government and foreign national government and commercial literature on semiconductors. *Journal of Business & Finance Librarianship* **29**, (2024).

[43] Dukes, S. & Bresniker, K. Systems and Architectures. *IEEE International Roadmap for Devices and Systems* 1–23 (2021) doi:https://doi.org/10.1109/irds54852.2021.00009

[44] Mustafa Badaroglu. More Moore. *IEEE International Roadmap for Devices and Systems* (2021) doi:https://doi.org/10.1109/irds54852.2021.00010.

[45] CMOS Logic IC STD (Standard) Series Outline, Application Note, Toshiba Electronic Devices & Storage Corporation, Jan. 2021, https://toshiba.semicon-storage.com/info/application_note_en_20210131_AKX00106.pdf?did=63520

[46] Barman, A. *et al.* Large electrocaloric effect in lead-free ferroelectric Ba0.85Ca0.15Ti0.9Zr0.1O3 thin film heterostructure. *APL Materials* **9**, (2021).

[47] Lin, Q., Wang, D. & Li, S. Strong Effect of Oxygen Partial Pressure on Electrical Properties of 0.5Ba(Zr$_{0.2}$Ti$_{0.8}$)O$_3$–0.5(Ba$_{0.7}$Ca$_{0.3}$)TiO$_3$ Thin Films. *Journal of the American Ceramic Society* **98**, 2094–2098 (2015).

[48] Chatterjee, S. *et al.* Interfacial strain induced giant magnetoresistance and magnetodielectric effects in multiferroic BCZT/LSMO thin film heterostructures. *Journal of Applied Physics* **135**, (2024).

[49] D. Hackenberg, R. Oldenburg, D. Molka, and R. Schone, *Introducing FIRESTARTER: A Processor Stress Test Utility*, Proc. Int. Green Computing Conference, (IGCC), 1-9, IEEE Press, 2013.

[50] V. Getov, D.J. Kerbyson, M. Macduff, and A. Hoisie, *Towards an Application-Specific Thermal Energy Model of Current Processors*, Proc. E2SC2015, pp. 1-10, ACM Press, 2015.

## Acknowledgements

Mohammad was funded by King Abdulaziz City for science and technology

PKP acknowledges the support from the Henry Royce Institute through the EPSRC grant EP/R00661X/1

## Author contributions statement

P.P. and V.G. conceived the project. M.A. and P.P. designed and conducted the procedures for thin film growth and output power estimation. V.G. conducted the experiment on the temperature of an Intel E5-2630 microprocessor as a function of power consumption. M.A., Q.Y. and C.U. conducted the experiment on X-ray photoelectron spectroscopy (XPS). All authors contributed to the analysis of the results and reviewed the manuscript.

## Data Availability

The data that support the findings of this study are not publicly available but are available from the corresponding author upon request.